\documentclass[12pt]{article}

\usepackage{tikz, adjustbox, float} 
\usepackage[round]{natbib} 
\usepackage{amsmath, amsfonts, amssymb, amsthm, dsfont} 
\usepackage{mathrsfs} 
\usepackage{soul} 
\usepackage{bm} 
\usepackage{booktabs}  
\usepackage{enumitem} 
\usepackage{mathabx} 
\usepackage{algorithm,algorithmic}

\def\real{{\rm I\!R}}

\DeclareMathOperator*{\argmin}{argmin}

\def\0{{\bf 0}}

\def\X{{\bf X}}
\def\x{{\bf x}}

\def\A{{\bf A}}

\DeclareMathOperator*{\var}{var}

\DeclareMathOperator*{\argzero}{argzero}

\usepackage{latexsym}
\usepackage{amsthm}
\usepackage{amsmath,amssymb,amsfonts,bm,amsbsy,bbm}
\usepackage{epsfig}
\usepackage{graphicx,rotating,lscape}
\usepackage{verbatim}
\usepackage{natbib}
\usepackage{multirow,color}
\usepackage{geometry}
\usepackage{setspace}
\usepackage{subfigure}
\usepackage{bbm}
\usepackage{ulem} 

\usepackage[para]{footmisc}

\def\X{{\bf X}}
\def\x{{\bf x}}

\def\u{{\bf u}}

\def\s{{\bf s}}
\def\a{{\bf a}}

\def\e{{\bf e}}

\def\bT{{\boldsymbol\Theta}}
\def\bb{{\boldsymbol\beta}}

\def\0{{\bf 0}}
\def\trans{^{\rm T}}
\def\pr{\hbox{pr}}
\def\wh{\widehat}
\def\wt{\widetilde}

\def\var{\hbox{var}}

\def\argmin{\mbox{argmin}}

\def\bse{\begin{eqnarray*}}
\def\ese{\end{eqnarray*}}
\def\be{\begin{eqnarray}}
\def\ee{\end{eqnarray}}
\def\bsq{\begin{equation*}}
\def\esq{\end{equation*}}
\def\bq{\begin{equation}}
\def\eq{\end{equation}}

\def\cS{\mathcal{S}}
\def\boxit#1{\vbox{\hrule\hbox{\vrule\kern6pt  \vbox{\kern6pt#1\kern6pt}\kern6pt\vrule}\hrule}}
\def\bse{\begin{eqnarray*}}
\def\ese{\end{eqnarray*}}
\def\be{\begin{eqnarray}}
\def\ee{\end{eqnarray}}
\def\bsq{\begin{equation*}}
\def\esq{\end{equation*}}
\def\bq{\begin{equation}}
\def\eq{\end{equation}}
\def\th{^{th}}
\def\var{\hbox{var}}

\def\wh{\widehat}
\def\wt{\widetilde}

\def\th{^{\rm th}}

\def\argmin{\mbox{argmin}}

\def\trans{^{\rm T}}

\def\bg{{\boldsymbol\gamma}}

\def\A{{\bf A}}
\def\a{{\bf a}}

\def\u{{\bf u}}

\def\u{{\bf u}}

\def\W{{\bf W}}

\def\X{{\bf X}}

\def\x{{\bf x}}

\def\bSig{{\bf \Sigma}}

\def\log{\hbox{log}}

\def\squarebox#1{\hbox to #1{\hfill\vbox to #1{\vfill}}}

\def\0{{\bf 0}}

\def\var{\hbox{var}}

\def\pr{\hbox{Pr}}
\def\wh{\widehat}
\def\wt{\widetilde}

\def\log{\hbox{log}}

\def\supp{\text{supp}}

\newtheoremstyle{mytheoremstyle} 
    {0.3cm}                      
    {0cm}                        
    {}                   
    {}                           
    {\bf}                   
    {: }                          
    {0em}                       
    {}  

\theoremstyle{mytheoremstyle}
\newtheorem{Theorem}{Theorem}

\newtheorem*{Lemma*}{Lemma}

\newtheorem{Proposition}{Proposition}

\newtheoremstyle{myExampleRemarkstyle} 
    {0.3cm}                    
    {0cm}                           
    {}                   
    {}                           
    {\bf}                   
    {: }                          
    {0em}                       
    {}  

\theoremstyle{myExampleRemarkstyle}
 
\newtheorem{Remark}{Remark}
\newtheorem{Assumption}{Assumption}

\newtheorem{innercustomgeneric}{\customgenericname}
\providecommand{\customgenericname}{}
\newcommand{\newcustomtheorem}[2]{%
  \newenvironment{#1}[1]
  {%
   \renewcommand\customgenericname{#2}%
   \renewcommand\theinnercustomgeneric{##1}%
   \innercustomgeneric
  }
  {\endinnercustomgeneric}
}

\newcustomtheorem{customthm}{Theorem}
\newcustomtheorem{customlemma}{Lemma}
\usepackage{mathalfa}
\usepackage{oubraces}

\let\refBKP\ref
\renewcommand{\ref}[1]{{\upshape\refBKP{#1}}}

\usepackage{setspace}
\begin{document}

\begingroup  
\begin{center}
\LARGE Accurate Inference for Penalized Logistic Regression 
\vspace{0.2cm}\\
\large{Yuming Zhang$^{1}$, St\'ephane Guerrier$^{2}$, Runze Li$^{3}$}
\vspace{0.2cm}\\
\small{$^{1}$Department of Biostatistics, Harvard University, USA \\ $^{2}$Faculty of Science, University of Geneva, Switzerland \\ $^{3}$Department of Statistics, Pennsylvania State University, USA}
\end{center}
\endgroup

\vspace{-0.2cm}  
\begin{abstract}

Inference for high-dimensional logistic regression models using penalized methods has been a challenging research problem. As an illustration, a major difficulty is the significant bias of the Lasso estimator, which limits its direct application in inference. Although various bias corrected Lasso estimators have been proposed, they often still exhibit substantial biases in finite samples, undermining their inference performance. These finite sample biases become particularly problematic in one-sided inference problems, such as one-sided hypothesis testing. This paper proposes a novel two-step procedure for accurate inference in high-dimensional logistic regression models. In the first step, we propose a Lasso-based variable selection method to select a suitable submodel of moderate size for subsequent inference. In the second step, we introduce a bias corrected estimator to fit the selected submodel. We demonstrate that the resulting estimator from this two-step procedure has a small bias order and enables accurate inference. Numerical studies and an analysis of alcohol consumption data are included, where our proposed method is compared to alternative approaches. Our results indicate that the proposed method exhibits significantly smaller biases than alternative methods in finite samples, thereby leading to improved inference performance.

\vspace{0.2cm}

\end{abstract}

\newpage
\section{Introduction}
\label{sec:intro}

The logistic regression model has been widely used to study binary data in various fields such as medicine, genetics, social sciences, and finance. It is assumed that the response $Y_i\in \{0,1\}$ with $i\in [n]\equiv \{1,\ldots,n\}$ is independently generated from 
\bq \label{eqn:logistic_model}
    Y_i | \x_i \sim \text{Bernoulli}\{g(\x_i\trans\bb^*)\} \quad \text{with} \quad g(x)\equiv \frac{\exp(x)}{1+\exp(x)},
\eq
where $\x_i$ is the associated deterministic $d$-dimensional explanatory covariate, and $\bb^*$ is the $d$-dimensional parameter. With the rapid growth of size of the data that are collected and processed, it becomes increasingly common to consider \textit{high-dimensional} data with $d> n$, preventing the direct application of classical methods such as the Maximum Likelihood Estimator (MLE) that are only applicable in \textit{low-dimensional} (i.e., $d$ is fixed and $d<n$) and some \textit{large-dimensional} settings (i.e., $d$ can diverge with $n$ but $d<n$). The extensive developments on penalized regression methods have made the analysis of high-dimensional data much more accessible (see e.g., \citealt{tibshirani1996regression,fan2001variable,zou2006adaptive,candes2007dantzig}), yet the inference still remains to be a challenging research question and has prompted different methodological developments.  

In this paper, we focus on inference for low-dimensional components in $\bb^*$. Since the construction of Confidence Intervals (CIs) and hypothesis testing are dual problems, our discussion will focus on the former. A major challenge encountered when performing inference with high-dimensional data is that many commonly used estimators are biased. A well known example is the Lasso \citep{tibshirani1996regression}, which suffers from severe biases that hinder its direct use for inference. Different bias corrected Lasso estimators have been proposed so that valid inference can be made (see e.g., \citealt{zhang2014confidence,van2014asymptotically,ning2017general}). Particularly for binary outcomes, \cite{ma2021global} proposed testing procedures for both the global null hypothesis and the large-scale simultaneous hypotheses for high-dimensional logistic regression models. \cite{guo2021inference} put forward a novel inference procedure for the case probability in a high-dimensional logistic regression model through the development of linearization and variance enhancement techniques. \cite{cai2023statistical} developed a unified inference framework for high-dimensional binary generalized linear models with general link functions. Overall, these methods in the debiasing Lasso literature treat the Lasso as a single procedure, deal with its bias correction, then make inference for the low-dimensional components of $\bb^*$. 

Despite the fact that these methods in the debiasing Lasso literature are asymptotically unbiased, they generally still remain considerably biased in finite samples that can lead to poor inference performance particularly for one-sided CIs. In fact, the impact of the bias arguably tends to be magnified when considering one-sided CIs compared to two-sided ones, where the effect of the bias may be (partially) compensated for the latter. The accurate construction of one-sided CIs is crucial to reliably address a wide range of research questions expressed as one-sided hypotheses across different fields. For example, in the medical field, consider a phase III confirmatory trial aiming to confirm the positive effect of an intervention (e.g., drug, treatment) compared to the reference group on a disease state (e.g., diseased or healthy) conditionally on individual characteristics (e.g., genetics, lifestyle), or a phase II trial investigating whether an improvement occurs after the use of a new treatment compared to the standard of care (see e.g., \citealt{boateng2019review} and the references therein). In social sciences, one-sided hypotheses are also often used. For example, in studies interested in determining the influence of certain individual traits (e.g., race, gender) on the occurrence of certain behaviors (e.g., alcohol consumption, drug abuse) or on certain decision makings, conditionally on other factors (e.g., education, family), assumptions are typically made regarding the direction of the effect related to the individual traits, like the effect of race on criminal sentencing, for example (see e.g., \citealt{steffensmeier1998interaction,fernando2006gender,kramer2020race}).

\subsection{Problem formulation}
\label{sec:intro:problem_formulation}

In this paper, we adopt a different approach than the debiasing Lasso literature. We consider a simple two-step approach, where we first adopt a model selection procedure, then we fit an estimator on the selected model for further inference. Such a two-step approach has been studied for high-dimensional linear regression (see e.g., \citealt{fan2012variance} for variance estimation and \citealt{zhao2021defense} for inference), yet it is less explored for logistic regression. 

In the context of high-dimensional logistic regression, one natural choice would be to first use the Lasso (or other model selection tools) to select a submodel of suitable and moderate size (smaller than the sample size), then to fit the MLE on the selected submodel. Naturally, one may expect this procedure to perform well if the selected submodel has the \textit{sure screening} property \citep{fan2008sure}, i.e., it contains the true model with probability approaching one and has a moderate model size smaller than $n$. Unfortunately, this procedure performs poorly in practice, mainly due to the following two problems. First, even when all significant variables are selected in the first step, the Lasso (and almost all other model selection methods) tends to additionally select irrelevant variables (i.e., with a true value of zero) that have large sample correlations with the response (i.e., spurious correlations). These spurious variables are selected to predict the realized noises, thereby causing the selected model to behave as if the true parameter values have been altered, even when all significant variables are included. For example, \cite{fan2012variance} illustrated the influence of spurious variables in introducing biases in variance estimation for high-dimensional linear regression. The second problem of this naive two-step procedure is that the MLE used to fit the selected submodel can be highly biased when the model size is relatively large compared to the sample size. In fact, even when there is no variable selection and we can fit the MLE on a large-dimensional full model, the MLE (despite being asymptotically unbiased) can suffer from severe finite sample biases which drastically impact the reliability of further inference. Indeed, \cite{sur2019modern} presented an empirical study to show that the MLE for a large-dimensional logistic regression model can be highly biased in finite samples, leading to subsequent inference problems such as the unreliable approximation for the distribution of the likelihood-ratio test. Therefore, in this paper we propose a two-step procedure that is able to mitigate the effects of these two problems, i.e., the issue caused by the spurious variables in the selected submodel and the large finite sample biases of the MLE in a large-dimensional selected submodel.

\subsection{Main results and contributions}
\label{sec:intro:contributions}

In this paper, we put forward a two-step procedure for inference of high-dimensional logistic regression models. In the first step, we propose a novel Lasso-based variable selection method to find a suitable submodel. In short, we split the data randomly into two halves, fit the Lasso on each half of the data for variable selection, then take the intersection of the variables selected based on each data subset as our final selected submodel. We call this variable selection method the \textit{Split Intersection LAsso} (SILA). In the second step, we fit a Bias Corrected MLE (BC-MLE) on the selected submodel. We refer to our proposed two-step procedure as the \textit{SILA with Bias correction} method, abbreviated as the SILAB.

Under suitable sparsity conditions, the submodel selected by the SILA enjoys the sure screening property. Moreover, the SILA method can significantly reduce the impact of irrelevant variables with high spurious correlations. An intuitive explanation is as follows: by splitting the data into halves, the sample size is reduced in each data subset, amplifying the influence of the spurious variables in each data subset. Then an irrelevant variable which has a large sample correlation with the response in one of the data subset will have a higher chance to be selected by the Lasso. Since the two data subsets are independent, the same irrelevant variable is unlikely to also have a large sample correlation with the response in the other data subset, and thus it is most likely not selected by the Lasso based on the other data subset. Therefore, when we take the intersection of the selected variables, irrelevant variables with high spurious correlations will tend to be discarded, thereby attenuating the influence of spurious variables.

The BC-MLE used in the second step can significantly reduce the biases of the MLE on the selected submodel under weak conditions. It is constructed based on the indirect inference proposed in \cite{gourieroux1993indirect} and following the approach used in \cite{guerrier2019simulation}. There are existing works (see e.g., \citealt{mackinnon1998approximate,gourieroux200013,guerrier2019simulation}) which have studied the bias correction properties of indirect inference based estimators. These works focus on bias correction for general parametric models and thus rely on higher-level assumptions, such as those on the unknown bias function, which are typically difficult to verify. In comparison, the BC-MLE is established specifically for the logistic regression models to reduce the bias of the MLE, and thus, its theoretical properties are established under a set of easily verifiable assumptions on the design. Moreover, we show that the BC-MLE has a smaller bias order than the MLE without loss of efficiency. The plug-in variance estimator based on the BC-MLE also has a smaller bias order than the one based on the MLE. The reduction of these biases allows to improve the accuracy of the inference performance using the BC-MLE than the MLE.

To summarize, the main contributions of this paper are as follows: 
\begin{itemize}
    \item Theoretically, our method only requires the selected submodel to have the sure screening property, thereby avoiding imposing stringent assumptions such as the irrepresentable condition or the beta-min condition that aim for model selection consistency (see e.g., \citealt{zhao2021defense}). Indeed, among all model consistent properties, the sure screening property is the weakest and the easiest to achieve in practice \citep{fan2012variance}. Our proposed SILA method for variable selection can search for a suitable model with the sure screening property while reducing the impact of the irrelevant variables with high spurious correlations. 
    
    \item We introduce the BC-MLE to fit on the selected submodel. The BC-MLE has a significantly smaller bias than the MLE in a large-dimensional model under weak conditions, which allows to improve the accuracy of subsequent inference.
    
    \item We present comprehensive numerical studies which demonstrate that our SILAB method can provide more accurate inference performance in finite samples than existing methods. In particular, when we construct one-sided CIs using our method, we observe empirical coverage probabilities generally closer to the nominal levels than alternative methods. When considering two-sided CIs, our method also provides accurate inference performance, but the corresponding improvement is less pronounced compared to that of alternative methods such as those in the debiasing Lasso literature. Therefore, our method can be seen as complementary to existing methods for high-dimensional inference in logistic regression models.
\end{itemize}

\subsection{Organization and notations}
\label{sec:intro:organization_notation}

The rest of the paper is organized as follows. In Section~\ref{sec:theory_submodel}, we derive and compare the asymptotic properties, particularly the bias orders, of both the MLE and the BC-MLE, as well as those of their plug-in variance estimators, conditional on a non-underfitted large-dimensional submodel. These results provide some insights into why the bias correction property of the BC-MLE can lead to improved inference performance. In Section~\ref{sec:HD_inference}, we present our proposed two-step procedure, the SILAB, for inference for high-dimensional logistic regression models. In Section~\ref{sec:simulations}, we present simulation studies to examine the validity of our theoretical findings as well as to demonstrate the advantageous inference performance of the SILAB method. A real data analysis on alcohol consumption is considered in Section~\ref{sec:case_study}, and lastly, we conclude the article and provide a discussion of future research directions in Section~\ref{sec:discussion}. The proofs of all theoretical results are collected in Supplementary Materials.

We finish this section by introducing some notations used throughout the paper. We denote $\supp(\bb)\equiv\{j:\beta_j\neq 0\}$. For any sets $\mathcal{A}$ and $\mathcal{B}$, we use $|\mathcal{A}|$ to denote the cardinality of $\mathcal{A}$, use $\mathcal{A}\subseteq\mathcal{B}$ to denote that $\mathcal{A}$ is a subset of $\mathcal{B}$, and use $\mathcal{A}\subset\mathcal{B}$ to denote that $\mathcal{A}$ is a strict subset of $\mathcal{B}$. For positive sequences $\{a_n\}$ and $\{b_n\}$, we write $a_n=o(b_n)$, $a_n\ll b_n$ or $b_n\gg a_n$ if $\lim_{n\to\infty} a_n/b_n = 0$. We write $a_n=\mathcal{O}(b_n)$ or $a_n \lesssim b_n$ if $a_n\leq C b_n$ for all $n$ with some finite positive constant $C$. We write $a_n\asymp b_n$ if $a_n = \mathcal{O}(b_n)$ and $b_n = \mathcal{O}(a_n)$ simultaneously. We denote $\overset{D}{\to}$ as convergence in distribution. In general, we use $C$ (or $c$) to denote some finite positive constants that are not necessarily the same. We denote $\e_{l,k}$ as a vector of length $l$ such that only the $k\th$ element is 1 and the rest are all 0. For a $m$-dimensional vector $\a$, we denote its $L_q$-norm as $\|\a\|_q \equiv (\sum_{i=1}^m |a_i|^q)^{1/q}$ with $1\leq q \leq
\infty$, and its $L_0$-norm $\|\a\|_0$ as the number of its non-zero elements. For a matrix $\A$, we write $A_{ij}$ as its $(i,j)\th$ component, and $\lambda_{\min}(\A),\lambda_{\max}(\A)$ as its minimum and maximum eigenvalues respectively.

\section{Methodology conditional on a non-underfitted submodel}
\label{sec:theory_submodel}

We write $\cS_F\equiv[d]$ and $\cS_T\equiv\supp(\bb^*)$ as the full and true models respectively. Suppose that there is an integer $d_0<n$ such that $|\cS_T|=d_0$, i.e., the original model in \eqref{eqn:logistic_model} has a sparsity level $d_0$. It is usually assumed that $d_0$ is fixed or diverges at a mild rate. Throughout this paper, we assume $d_0^2\log(n)n^{-1}\to 0$ as $n\to\infty$. Moreover, suppose that we are interested in inference for the low-dimensional components of $\bb^*$ indexed by $\cS_I$, where $\cS_I\subset\cS_F$ and $|\cS_I|<n$. Without loss of generality, we consider $\cS_I = \{j_0\}$ for a given $j_0\in\cS_F$.

Consider a \textit{deterministic} submodel $\cS\subset\cS_F$ of dimension $p$ with
\bq \label{eqn:logistic_submodel}
    Y_i|\x_{i,\cS} \sim \text{Bernoulli}\{g(\x_{i,\cS}\trans\bb_{\cS})\},
\eq
where $\x_{i,\cS}$ denote the associated covariate vector for the $i\th$ subject and $\bb_{\cS}$ denotes the true regression coefficient vector given the considered submodel. Let $\bb_\cS^*$ be the subvector containing the components of $\bb^*$ (i.e., the true parameter of the original model in \eqref{eqn:logistic_model}) that are indexed by $\cS$. In general, $\bb_{\cS} \neq \bb_\cS^*$, but when $\cS_T \subseteq \cS$, we have $\bb_{\cS} = \bb_\cS^*$. In this section, we consider a submodel $\cS$ that is non-underfitted, contains our parameters of interest, has a moderate size smaller than $n$, and has a regular parameter space, as described in Assumption~\ref{assume:submodel}. 

\begin{Assumption}
\label{assume:submodel}
The deterministic submodel $\cS$ satisfies: (i) $\cS_T \subseteq \cS$; (ii) $\cS_I \subseteq \cS$; (iii) $p^2\log(n)n^{-1}\to 0$ as $n\to\infty$; (iv) the parameter space of $\bb_{\cS}$, denoted as $\bT_{\cS}$, is a compact convex subset of $\real^{p}$, and $\bb_{\cS}$ lies in the interior of $\bT_{\cS}$. 
\end{Assumption}

A direct consequence of Assumption~\ref{assume:submodel} is that $\bb_{\cS} = \bb_\cS^*$. Hence, the observed data $\{Y_i\}_{i\in[n]}$ generated from the original model in \eqref{eqn:logistic_model} based on $\{\x_i\}_{i\in[n]}$ and $\bb^*$ can be viewed as being generated from the submodel in \eqref{eqn:logistic_submodel} based on $\{\x_{i,\cS}\}_{i\in[n]}$ and $\bb_\cS$. Based on the submodel $\cS$, we can compute the MLE for $\bb_\cS$ (and equivalently for $\bb_\cS^*$) as
\bq \label{eqn:def_MLE}
    \wt{\bb}_\cS \equiv \argzero_{\bg\in\bT_{\cS}} \sum_{i=1}^n \left\{Y_i - g(\x_{i,\cS}\trans\bg)\right\} \x_{i,\cS}.
\eq
As illustrated in Section~\ref{sec:intro}, despite being asymptotically unbiased under regularity conditions, the MLE may suffer from severe finite sample biases when $p$ (the dimension of $\cS$) is relatively large compared to $n$. Thus, we propose to construct the BC-MLE as follows:
\bq \label{eqn:def_JINI}
    \wh{\bb}_\cS \equiv \argzero_{\bg\in\bT_{\cS}} \wt{\bb}_{\cS} - \mathbb{E}_{\bg}\left\{\wt{\bb}_{\cS}(\bg)\right\},
\eq
where $\wt{\bb}_{\cS}(\bg)$ denotes the value of the MLE $\wt{\bb}_\cS$ computed on a sample of size $n$ generated from the submodel \eqref{eqn:logistic_submodel} based on $\{\x_{i,\cS}\}_{i\in[n]}$ and $\bg$, and $\mathbb{E}_\bg(\cdot)$ denotes the expectation with respect to the submodel in \eqref{eqn:logistic_submodel} based on $\{\x_{i,\cS}\}_{i\in[n]}$ and $\bg$. More details on the numerical implementation to solve \eqref{eqn:def_JINI} can be found in Section~\ref{sec:HD_inference:psi}.

Next, we demonstrate that, under suitable conditions, the BC-MLE is asymptotically optimal in the sense that it has the same limiting distribution as the MLE. We define $\bSig_n(\bg)\equiv \X_\cS\trans\W_\cS(\bg)\X_\cS$ for $\bg\in\bT_\cS$, where $\X_\cS$ is the $n\times p$ design matrix with the $i\th$ row as $\x_{i,\cS}\trans$, and $\W_\cS(\bg)$ is the $n\times n$ diagonal matrix with the $i\th$ diagonal entry as $\var_\bg(Y_i|\x_{i,\cS})$. Under regularity conditions, this matrix is the inverse of the covariance matrix of the MLE for a logistic regression based on the design $\X_\cS$ and the parameter $\bg$. In order to study the asymptotic normality of the BC-MLE, we consider suitable conditions on the design as specified in Assumption~\ref{assume:submodel:covariate} below.

\begin{Assumption}
\label{assume:submodel:covariate}
The design matrix $\X_\cS$ satisfies: (i) $\sum_{i=1}^n \|\x_{i,\cS}\|_2^4=\mathcal{O}(np^2)$; \\ (ii) $\sup_{\u\in\real^{p}:\|\u\|_2=1} \sum_{i=1}^n (\u\trans\x_{i,\cS})^4 = \mathcal{O}(n)$; (iii) $\inf_{\u\in\real^{p}:\|\u\|_2=1} \sum_{i=1}^n (\u\trans\x_{i,\cS})^2 \geq cn$ with large enough $n$; (iv) $\min_{i\in[n]}\|\x_{i,\cS}\|_2 > 0$; (v) $\min_{i\in[n]}\inf_{\bg\in\bT_{\cS}} \var_\bg(Y_i|\x_{i,\cS})\geq c>0$. 
\end{Assumption}

Assumption~\ref{assume:submodel:covariate} consists of standard regularity conditions on the design. For example, Condition (i) can be satisfied when $\max_{i\in[n]}\|\x_{i,\cS}\|_2=\mathcal{O}(p^{1/2})$. This is the case, for example, when the covariates $\{\x_{i,\cS}\}_{i\in[n]}$ are standardized and lie in a compact space $[-c,c]^p$. Moreover, Conditions (ii) and (iii) can be satisfied when $c\leq \min_{i\in[n]}\inf_{\u\in\real^p:\|\u\|_2=1} |\u\trans\x_{i,\cS}| \leq \max_{i\in[n]}\sup_{\u\in\real^p:\|\s\|_2=1} |\u\trans\x_{i,\cS}| \leq C$. A more commonly used alternative in the literature would be $cn\leq \lambda_{\min}(\X_{\cS}\trans\X_{\cS}) \leq \lambda_{\max}(\X_{\cS}\trans\X_{\cS}) \leq Cn$. Condition (iv) ensures that all $\{\x_{i,\cS}\}_{i\in[n]}$ vectors are not trivially equal to zero. Condition (v) ensures that the variance of the response variable from a logistic regression based on $\x_{i,\cS}$ and any parameter from $\bT_{\cS}$ is bounded below and away from zero for all $i\in [n]$.

We can now establish the asymptotic normality of the BC-MLE as in Proposition~\ref{result:submodel:asymp_norm} below.

\begin{Proposition}
\label{result:submodel:asymp_norm}
    Consider a deterministic submodel $\cS$ which satisfies Assumption~\ref{assume:submodel}. Then under Assumption~\ref{assume:submodel:covariate}, for any $\u\in\real^p$ such that $\|\u\|_2=1$, we have
    \bsq
        \frac{\sqrt{n}\u\trans(\wt{\bb}_{\cS} - \bb_{\cS}^*)}{\sigma_{n,\u}(\bb_\cS^*)} \overset{D}{\to} \mathcal{N}(0,1) \quad \text{and} \quad \frac{\sqrt{n}\u\trans(\wh{\bb}_{\cS} - \bb_{\cS}^*)}{\sigma_{n,\u}(\bb_\cS^*)} \overset{D}{\to} \mathcal{N}(0,1),
    \esq
    where $\sigma_{n,\u}^2(\bb_\cS^*)\equiv n\u\trans\bSig_n(\bb_\cS^*)^{-1}\u$ and $\sigma_{n,\u}(\bb_\cS^*)\equiv \sqrt{\sigma_{n,\u}^2(\bb_\cS^*)}$. 
\end{Proposition}

Proposition~\ref{result:submodel:asymp_norm} shows that the MLE and the BC-MLE have the same limiting distribution with respect to the same population target $\bb_\cS^*$. Based on this result, we can construct asymptotically valid CIs for any linear combination $\u\trans\bb_\cS^*$. As an illustration, the $(1-\alpha)$ level two-sided CIs based on the BC-MLE can be constructed as 
\bsq
    \text{CI}_{\alpha}^{(\cS)}(\wh{\bb}_\cS;\u\trans\bb_{\cS}^*) = \u\trans\wh{\bb}_\cS \pm z_{1-\alpha/2} \frac{\sigma_{n,\u}(\wh{\bb}_\cS)}{\sqrt{n}},
\esq
where $z_{1-\alpha/2}$ denotes the $(1-\alpha/2)$ quantile of a standard normal distribution. The CIs based on the MLE can be constructed in the same way by plugging in $\wt{\bb}_\cS$. Particularly, we let $\u=\e_{p,j_0'}$ where $j_0'\in[p]$ denotes the index such that $\beta_{\cS,j_0'}^*$ corresponds to $\beta_{j_0}^*$, then CIs can be constructed for $\beta_{j_0}^*$. It is reasonable to expect that more accurate approximations to the expected CI center and standard deviation can improve the accuracy of CIs. Therefore, we study below the bias orders of the MLE and the BC-MLE (see Proposition~\ref{result:submodel:bias_point_est}), as well as the ones of their plug-in variance estimators (see Proposition~\ref{result:submodel:bias_plugin_var}). We show that the BC-MLE and its plug-in variance estimator have smaller bias orders than the ones of the MLE, thereby allowing for more accurate inference. 

\begin{Proposition}
\label{result:submodel:bias_point_est}
    Consider a deterministic submodel $\cS$ which satisfies Assumption~\ref{assume:submodel}. Then under Assumption~\ref{assume:submodel:covariate}, for any $\u\in\real^p$ such that $\|\u\|_2=1$, we have
    \bsq
        \sup_{\u\in\real^p:\|\u\|_2=1} \left|\mathbb{E}(\u\trans\wt{\bb}_{\cS}) - \u\trans\bb_{\cS}^*\right| = \mathcal{O}(pn^{-1}) \;\; \text{and} \;\; \sup_{\u\in\real^p:\|\u\|_2=1} \left|\mathbb{E}(\u\trans\wh{\bb}_{\cS}) - \u\trans\bb_{\cS}^*\right| = \mathcal{O}(p^{3/2}n^{-3/2}).
    \esq
\end{Proposition}

We can interpret these orders by considering various diverging rates of $p$, noting that $p\ll n^{1/2}$ by Assumption~\ref{assume:submodel}. When $p$ is fixed, Proposition~\ref{result:submodel:bias_point_est} shows that the MLE has a bias order of $\mathcal{O}(n^{-1})$ which is in line with existing results (see e.g., \citealt{cox1968general,cox1979theoretical}). On the other hand, the BC-MLE has a bias order of $\mathcal{O}(n^{-3/2})$, with an improvement of $\mathcal{O}(n^{-1/2})$ compared to the bias order of the MLE. When $p \lesssim n^{1/3}$, the bias of the MLE is $\mathcal{O}(n^{-2/3})$ and the one of the BC-MLE is $\mathcal{O}(n^{-1})$ with an improvement of $\mathcal{O}(n^{-1/3})$. When $n^{1/3}\ll p \ll n^{1/2}$, the bias of the MLE is $o(n^{-1/2})$ and the one of the BC-MLE is $o(n^{-3/4})$ with an improvement of $\mathcal{O}(n^{-1/4})$. Therefore, overall the BC-MLE has a smaller bias order than the MLE, rendering a more accurate approximation to the expected CI center. 

We consider an additional assumption on the design which is only used to study the bias orders of the plug-in variance estimators. 

\begin{Assumption}
\label{assume:submodel:covariate2}
    The design matrix $\X_\cS$ satisfies $\sup_{\u\in\real^{p}: \|\u\|_2=1} \sum_{i=1}^n (\u\trans\x_{i,\cS})^8 = \mathcal{O}(n)$. 
\end{Assumption}

Assumption~\ref{assume:submodel:covariate2} is a plausible requirement on the design, which can be satisfied, for example, when $\max_{i\in[n]}\sup_{\u\in\real^{p}:\|\u\|_2=1}|\u\trans\x_{i,\cS}|<C$.

We present in Proposition~\ref{result:submodel:bias_plugin_var} below the bias orders of the plug-in variance estimators based on the MLE $\wt{\bb}_{\cS}$ and the BC-MLE $\wh{\bb}_{\cS}$. 

\begin{Proposition}
\label{result:submodel:bias_plugin_var}
    Consider a deterministic submodel $\cS$ which satisfies Assumption~\ref{assume:submodel}. Then under Assumptions~\ref{assume:submodel:covariate} and \ref{assume:submodel:covariate2}, for any $\u\in\real^{p}$ such that $\|\u\|_2=1$, we have 
    \bse
        && \mathbb{E}\{\sigma_{n,\u}^2(\wt{\bb}_{\cS})\} - \sigma_{n,\u}^2(\bb_{\cS}^*) = \min\{\mathcal{O}(p^{3/2}n^{-1}),o(n^{-1/2})\}, \\
        && \mathbb{E}\{\sigma_{n,\u}^2(\wh{\bb}_{\cS})\} - \sigma_{n,\u}^2(\bb_{\cS}^*) = \max\{\mathcal{O}(p^2n^{-3/2}),o(p^{1/2}n^{-1})\}.
    \ese
\end{Proposition}

Similarly to the discussion for Proposition~\ref{result:submodel:bias_point_est}, we can compare these bias orders by considering different diverging rates of $p$. When $p$ is fixed, the bias order of the plug-in variance estimator based on the MLE is $\mathcal{O}(n^{-1})$, whereas the one based on the BC-MLE is $o(n^{-1})$ with an improvement of $o(1)$. When $p \lesssim n^{1/3}$, the bias order of the one based on the MLE is $o(n^{-1/2})$, whereas the one based on the BC-MLE is $\mathcal{O}(n^{-5/6})$ with an improvement of $\mathcal{O}(n^{-1/3})$. When $n^{1/3}\ll p \ll n^{1/2}$, the bias orders of both plug-in variance estimators are $o(n^{-1/2})$. Therefore, the plug-in variance estimator based on the BC-MLE generally provides a less biased estimation of the asymptotic variance than the one based on the MLE. 

Since the BC-MLE provides less biased estimators to both $\u\trans\bb_{\cS}^*$ (Proposition~\ref{result:submodel:bias_point_est}) and $\sigma_{n,\u}^2(\bb_{\cS}^*)$ (Proposition~\ref{result:submodel:bias_plugin_var}), it is reasonable to believe that it can lead to the construction of CIs with more accurate empirical coverage probabilities. This intuition is supported by a simulation study presented in Supplementary Materials. In this numerical experiment, we consider a logistic regression with $d<n$ so that we can use the full model as our considered submodel, i.e., let $\cS=\cS_F$ and thus $p=d$. We compare the finite sample performance of the MLE and the BC-MLE in terms of estimation and inference for a single coefficient $\beta_4^*$ (i.e., the fourth component of $\bb^*$) with a true value of $0.25$. We consider three settings with diverging $p$ and $n$ so that the asymptotic behaviors of both estimators can be assessed. Our numerical results show the following: (i) The BC-MLE appears to be unbiased in estimating $\beta_4^*$ even when $p$ is relatively large compared to $n$, whereas the MLE is significantly biased especially in large $p/n$ settings. (ii) The plug-in variance estimator based on the BC-MLE has a negligible bias even when $p/n$ is large, whereas the one based on the MLE is biased. (iii) When considering the one-sided test $\text{H}_0:\beta_4^*=0.25 \;\text{vs}\; \text{H}_1:\beta_4^*>0.25$, the BC-MLE leads to an empirical type-I error that is generally closer to the nominal level than the MLE. (iv) When considering the one-sided test $\text{H}_0:\beta_4^*=0 \;\text{vs}\; \text{H}_1:\beta_4^*> 0$, both estimators lead to comparable statistical power. Therefore, these numerical results support our theoretical findings and emphasize that the higher-order bias correction property of the BC-MLE can significantly improve the inference accuracy in finite samples, particularly when considering one-sided inference problems.

\section{Accurate inference for high-dimensional logistic regression}
\label{sec:HD_inference}

Section~\ref{sec:theory_submodel} provides the theoretical guarantees of the BC-MLE after simplifying the high-dimensional inference problem on the original model of \eqref{eqn:logistic_model} to a large-dimensional one through a non-underfitted submodel $\cS$. In this section, we present the novel SILA method which allows to find such a suitable submodel (see Section~\ref{sec:HD_inference:selection}). Based on this selected submodel, asymptotically valid and accurate inference can be made using the BC-MLE (see Section~\ref{sec:HD_inference:psi}). These two steps constitute our proposed SILAB method.

\subsection{Split intersection lasso} 
\label{sec:HD_inference:selection}

We propose the SILA method which allows us to (i) obtain a submodel $\wh{\cS}$ of moderate size that does not underfit asymptotically, i.e., with a sure screening property; (ii) reduce the impact of irrelevant variables with high spurious correlations. The SILA method consists of several steps, and for clarity and exposition, we present it in Algorithm~\ref{algo:selection}. 

\begin{algorithm}[tb]
\caption{The Split Intersection LAsso (SILA) method}
\label{algo:selection}
\begin{algorithmic}[1]
\renewcommand{\algorithmicrequire}{\textbf{Input:}}
\renewcommand{\algorithmicensure}{\textbf{Output:}}
\REQUIRE observed data $(Y_i,\x_i)_{i\in[n]}$, non-negative tuning parameters $\delta_1,\delta_2$
\vspace{0.1cm} \ENSURE an estimated submodel $\wh{\cS}$
\\ \vspace{0.1cm} \STATE Randomly and evenly split the observed data into two sets, say $(Y_i,\x_i)_{i\in\mathcal{I}_1}$ and $(Y_i,\x_i)_{i\in\mathcal{I}_2}$, where $\mathcal{I}_1\cup\mathcal{I}_2=[n]$, $\mathcal{I}_1\cap\mathcal{I}_2=\varnothing$ and $|\mathcal{I}_1| = |\mathcal{I}_2|$.
\\ \vspace{0.1cm} \STATE Consider a strictly increasing sequence of regularization parameters of the Lasso $\Lambda(\delta_1,\delta_2) \equiv \{\lambda_1,\ldots,\lambda_K\}$ in $[0,+\infty)$ such that 
\bsq
    \delta_1 < \min\left\{\left\|\wh{\bb}_{\lambda_K}^{(1)}\right\|_0,\left\|\wh{\bb}_{\lambda_K}^{(2)}\right\|_0\right\} \leq \max\left\{\left\|\wh{\bb}_{\lambda_1}^{(1)}\right\|_0,\left\|\wh{\bb}_{\lambda_1}^{(2)}\right\|_0\right\} < \delta_2,
\esq
with the Lasso estimators
\bsq
    \wh{\bb}_{\lambda_k}^{(q)} \equiv \underset{\bb\in\real^d}{\argmin} \left\{-2\sum_{i\in\mathcal{I}_q} l(y_i, \x_i\trans\bb) + \lambda_k \|\bb\|_1 \right\} \quad \text{for} \quad k\in[K], \; q\in\{1,2\},
\esq
where $l(y_i, \x_i\trans\bb)\equiv y_i\log\{g(\x_i\trans\bb)\} + (1-y_i)\log\{1-g(\x_i\trans\bb)\}$ denotes the log-likelihood. 
\\ \vspace{0.1cm} \STATE Compute 
\bq \label{eqn:lambda_AIC}
    \wh{\lambda}_q \equiv \underset{\lambda\in\Lambda(\delta_1,\delta_2)}{\argmin} \left[-2\sum_{i\in\mathcal{I}_q} l\left\{y_i, \x_i\trans\wh{\bb}_{\lambda}^{(q)}\right\} + \left\|\wh{\bb}_{\lambda}^{(q)}\right\|_0\right] \quad \text{for} \quad q\in\{1,2\}.
\eq
\\ \STATE Let $\wh{\cS}^{(q)} \equiv \supp\{\wh{\bb}_{\wh{\lambda}_q}^{(q)}\}$ for $q\in\{1,2\}$, output $\wh{\cS} \equiv \{\wh{\cS}^{(1)} \cap \wh{\cS}^{(2)}\} \cup \{j_0\}$.
\end{algorithmic} 
\end{algorithm}

In Steps~1 and 2 in Algorithm~\ref{algo:selection}, we focus on regularization parameters $\lambda$ used in the Lasso whose corresponding models have sizes determined by $\delta_1,\delta_2$. Then we compute the Lasso on two data subsets after data splitting. Since each data subset only has a reduced sample size, the dimensionality in each data subset becomes larger, hence enhancing the sample correlations of the irrelevant variables with the response. In Step~3, we find the optimal regularization parameters $\wh{\lambda}_1,\wh{\lambda}_2$ for both data subsets based on an Akaike Information Criterion (AIC) \citep{akaike1974new}, where the penalty constant in \eqref{eqn:lambda_AIC} is reduced to one instead of the standard value of two to account for the half sample size $n/2$ of each data subset. Such a choice of the variable selection criterion ensures that the probability of underfitting goes to zero with a tendency of overfitting \citep{zhang2010regularization}. As a consequence, $\wh{\cS}^{(1)}$ and $\wh{\cS}^{(2)}$, the models selected based on both data subsets, contain the true model $\cS_T$ with probability approaching one. Lastly, in Step~4 we take the intersection of these two selected models as our final $\wh{\cS}$. We enforce the parameter of interest $j_0$ to be included so that further inference can be made for $\beta_{j_0}^*$ based on $\wh{\cS}$. As mentioned in Section~\ref{sec:intro}, an irrelevant variable with a high spurious correlation is unlikely to be selected in both $\wh{\cS}^{(1)}$ and $\wh{\cS}^{(2)}$. Thus, taking the intersection is the key step to discard most of the irrelevant variables with high spurious correlations, allowing to reduce the impact of spurious variables. Moreover, since both $\wh{\cS}^{(1)}$ and $\wh{\cS}^{(2)}$ contain $\cS_T$ with probability approaching one, so does their intersection $\wh{\cS}$. Therefore, the selected submodel $\wh{\cS}$ by the SILA method is guaranteed to have a sure screening property with reduced influence of the spurious variables. 

\begin{Remark}
    The inclusion of $j_0$ in the final selected submodel $\wh{\cS}$ can also be achieved by putting no penalty on the corresponding component in the Lasso in Step~2. Then we guarantee $\wh{\cS}^{(1)}$ and $\wh{\cS}^{(2)}$ to include $j_0$, and can simply output $\wh{\cS} \equiv \{\wh{\cS}^{(1)} \cap \wh{\cS}^{(2)}\}$ in Step~4.
\end{Remark}

\begin{Remark}
    As mentioned in Section~\ref{sec:theory_submodel}, we consider the parameter of interest to be $\beta_{j_0}^*$ for simplicity of discussion, and thus in Algorithm~\ref{algo:selection} we only enforce the inclusion of $j_0$. Naturally, the SILA method can be extended to include a set of low-dimensional components of $\bb^*$, which can then be used to perform inference, for example, on a linear combination of these low-dimensional components of $\bb^*$. 
\end{Remark}

A suitable choice of the tuning parameters $\delta_1,\delta_2$ is the key to ensure that the selected submodel $\wh{\cS}$ has a moderate size and asymptotically valid inference for $\beta_{j_0}^*$ can be made based on $\wh{\cS}$. As an illustration, we denote $\cS_{\cup}$ as the set of all submodels that satisfy Assumption~\ref{assume:submodel}, i.e., $\cS_{\cup} \equiv \{\cS: {\text{$\cS$ satisfies Assumption~\ref{assume:submodel}}}\}$. Then the coverage probability of the $(1-\alpha)$ level two-sided CI for $\beta_{j_0}^*$ based on $\wh{\cS}$ satisfies
\bse
    \pr\left\{\beta_{j_0}^* \in \text{CI}_\alpha^{(\wh{\cS})}(\wh{\bb}_{\wh{\cS}}; \beta_{j_0}^*)\right\} &=& \pr\left\{\beta_{j_0}^* \in \text{CI}_\alpha^{(\wh{\cS})}(\wh{\bb}_{\wh{\cS}}; \beta_{j_0}^*) \Big| \wh{\cS}\in \cS_{\cup}\right\} \pr(\wh{\cS} \in \cS_{\cup}) \\
    &&+ \pr\left\{\beta_{j_0}^* \in \text{CI}_\alpha^{(\wh{\cS})}(\wh{\bb}_{\wh{\cS}}; \beta_{j_0}^*) \Big| \wh{\cS} \notin \cS_{\cup}\right\} \pr(\wh{\cS} \notin \cS_{\cup}) \\
    &\to& 1-\alpha \quad \text{as} \quad n\to\infty,
\ese 
under Assumption~\ref{assume:submodel:selection} on $\delta_1,\delta_2$ given as follows.

\begin{Assumption}
\label{assume:submodel:selection}
The non-negative tuning parameters $\delta_1,\delta_2$ in Algorithm~\ref{algo:selection} are such that $\lim_{n\to\infty} \pr(\wh{\cS} \in \cS_{\cup}) = 1$, i.e., (i) $\wh{\cS}$ does not underfit asymptotically, (ii) the size of $\wh{\cS}$ satisfies $|\wh{\cS}|^2\log(n)n^{-1} \to 0$, and (iii) the parameter space of $\bb_{\wh{\cS}}^*$ is compact and convex.  
\end{Assumption}

Condition (i) can be trivially satisfied even with $\delta_1=0$ and $\delta_2=+\infty$. Indeed, since we use a bounded positive constant as the penalty for $\|\wh{\bb}_\lambda^{(q)}\|_0$ in \eqref{eqn:lambda_AIC} for the selection of regularization parameters, the corresponding models $\wh{\cS}^{(1)}$ and $\wh{\cS}^{(2)}$ selected based on the data subsets do not underfit and tend to overfit asymptotically based on the results of \cite{zhang2010regularization}. Since $\wh{\cS}$ takes the intersection of $\wh{\cS}^{(1)}$ and $\wh{\cS}^{(2)}$, it is also guaranteed to not underfit asymptotically and thus satisfying Condition (i). Condition (ii) ensures that the selected model $\wh{\cS}$ is of moderate size given the choices of $\delta_1,\delta_2$, and provides insights into the relation between $\delta_1,\delta_2$ and $d,n$, where we recall that $d$ is the dimension of the full model $\cS_F$. In practice, we propose to use $\delta_1 = n/12$ and $\delta_2 = n/2$ in relatively balanced settings (i.e., when $\sum_{i= 1}^n Y_i \approx n/2$), and in unbalanced settings we recommend to use $\delta_1 =\min(\sum_{i= 1}^n Y_i, n -\sum_{i= 1}^n Y_i)/6$. This choice is mainly based on our empirical experience and a desire for simplicity. It also relies on a crude approximation considering $d \gg n^{3/2}$. Indeed, suppose that the model is null (i.e., the sparsity level $d_0=0$) and that all $d$ number of irrelevant variables are equally likely to be selected based on each data subset in Step~2 of Algorithm~\ref{algo:selection}. Then the number of irrelevant variables selected based on both data subsets would follow a Binomial distribution $B(d, \frac{\|\wh{\bb}_\lambda^{(1)}\|_0\|\wh{\bb}_\lambda^{(2)}\|_0}{d^2})$. In this case, our suggested choices of $\delta_1,\delta_2$ can satisfy Condition (ii) when $d \gg n^{3/2}$. Finally, Condition (iii) is a standard regularity condition on the parameter space.

\subsection{The SILA with bias correction method}
\label{sec:HD_inference:psi}

Based on the submodel selected by the SILA method, we can fit the BC-MLE for further inference. This SILAB procedure is provided in Algorithm~\ref{algo:psi}, where we define the final estimator $\wh{\bb}_{\wh{\cS}}$ for $\bb_{\wh{\cS}}^*$ and $\wh{\beta}_{j_0}$ for $\beta_{j_0}^*$. These estimators generated through the SILAB procedure are referred to as the SILAB estimators.

\begin{Remark}
    A convenient algorithm to compute the BC-MLE is the Iterative Bootstrap (IB) algorithm proposed in \cite{kuk1995asymptotically}. \cite{guerrier2019simulation} showed that the limit of the IB algorithm is equivalent to the simulation-based approximated version of the BC-MLE. In Algorithm~\ref{algo:psi}, we use the IB algorithm to compute the BC-MLE on the submodel selected by the SILA method. 
\end{Remark}

\begin{algorithm}[tb]
\caption{The SILA with Bias correction (SILAB) method}
\label{algo:psi}
\begin{algorithmic}[1]
\renewcommand{\algorithmicrequire}{\textbf{Input:}}
\renewcommand{\algorithmicensure}{\textbf{Output:}}
\REQUIRE observed data $(Y_i,\x_i)_{i\in[n]}$, non-negative tuning parameters $\delta_1,\delta_2$, tolerance level $\epsilon$, number of simulated samples $H$
\vspace{0.1cm} \ENSURE the SILAB estimator $\wh{\beta}_{j_0}$ and its asymptotic variance estimator $\wh{\sigma}_{j_0}^2$
\\ \vspace{0.1cm} \STATE Compute $\wh{\cS}$ using the SILA method provided in Algorithm~\ref{algo:selection}. 
\\ \vspace{0.1cm} \STATE Given $\wh{\cS}$, compute the MLE $\wt{\bb}_{\wh{\cS}}$ as given in \eqref{eqn:def_MLE} using the data $(Y_i,\x_{i,\wh{\cS}})_{i\in[n]}$. 
\\ \vspace{0.1cm} \STATE Let $\wh{\bb}_{\wh{\cS}}^{(0)} \equiv \wt{\bb}_{\wh{\cS}}$. Initiate $k=0$ and $\epsilon_0=\epsilon$.
\\ \vspace{0.1cm} 
\WHILE {$\epsilon_k \geq \epsilon$}
\vspace{0.1cm} \STATE Let $k = k+1$. 
\vspace{0.1cm} \STATE Simulate $H$ independent samples of size $n$ denoted as $\{Y_{i}^{(h)}\}_{i\in[n]}$ with $h\in[H]$ from the submodel \eqref{eqn:logistic_submodel} using covariates $(\x_{i,\wh{\cS}})_{i\in[n]}$ and $\wh{\bb}_{\wh{\cS}}^{(k-1)}$. 
\vspace{0.1cm} \STATE For each $h\in[H]$, compute the MLE on the data $\{Y_i^{(h)}, \x_{i,\wh{\cS}}\}_{i\in[n]}$ and denote it as $\wt{\bb}_{\wh{\cS}}^{(k-1,h)}$.
\vspace{0.1cm} \STATE Compute $$\wh{\bb}_{\wh{\cS}}^{(k)} \equiv \wh{\bb}_{\wh{\cS}}^{(0)} + \wh{\bb}_{\wh{\cS}}^{(k-1)} - \frac{1}{H}\sum_{h=1}^H \wt{\bb}_{\wh{\cS}}^{(k-1,h)},$$ and $\epsilon_k \equiv \|\wh{\bb}_{\wh{\cS}}^{(k)} - \wh{\bb}_{\wh{\cS}}^{(k-1)}\|_2$. 
\vspace{0.1cm} \ENDWHILE
\vspace{0.1cm} \STATE Let $\wh{\bb}_{\wh{\cS}} \equiv \wh{\bb}_{\wh{\cS}}^{(k)}$, which is the SILAB estimator for $\bb_{\wh{\cS}}^*$.  
\\ \vspace{0.1cm} \STATE Let $j_0'\in[|\wh{\cS}|]$ denote the index such that $\beta_{\wh{\cS},j_0'}^*$ corresponds to $\beta_{j_0}^*$, then we set $\wh{\beta}_{j_0} \equiv \wh{\beta}_{\wh{\cS},j_0'}$ as the SILAB estimator for $\beta_{j_0}^*$ and $\wh{\sigma}_{j_0}^2 \equiv n\{\bSig_n(\wh{\bb}_{\wh{\cS}})^{-1}\}_{(j_0',j_0')}$ as the plug-in variance estimator for $\sigma_{j_0}^2 \equiv n\{\bSig_n(\bb^*_{\wh{\cS}})^{-1}\}_{(j_0',j_0')}$.
\end{algorithmic} 
\end{algorithm} 

Exploiting the results in Sections~\ref{sec:theory_submodel} and \ref{sec:HD_inference:selection}, the theoretical properties of the SILAB estimator $\wh{\beta}_{j_0}$ and its plug-in variance estimator $\wh{\sigma}_{j_0}^2$ can be summarized below. 

\begin{Theorem}
\label{result:psi:asymp_norm}
    Consider the SILAB estimator $\wh{\beta}_{j_0}$ for $\beta_{j_0}^*$ and its plug-in variance estimator $\wh{\sigma}_{j_0}^2$ obtained from Algorithm~\ref{algo:psi}. Under Assumptions~\ref{assume:submodel:covariate} and \ref{assume:submodel:selection}, we have 
    \bsq
        \frac{\sqrt{n}(\wh{\beta}_{j_0}-\beta_{j_0}^*)}{\wh{\sigma}_{j_0}} \overset{D}{\to} \mathcal{N}(0,1), 
    \esq
    with $\wh{\sigma}_{j_0} \equiv \sqrt{\wh{\sigma}_{j_0}^2}$.
\end{Theorem}

Theorem~\ref{result:psi:asymp_norm} states that our proposed SILAB method can provide asymptotically valid inference for $\beta_{j_0}^*$. This result allows, among others, to approximate a $(1-\alpha)$ level two-sided CI for $\beta_{j_0}^*$ with $\wh{\beta}_{j_0} \pm z_{1-\alpha/2} n^{-1/2} \wh{\sigma}_{j_0}$.  

\begin{Theorem}
\label{result:psi:bias}
    Consider the SILAB estimator $\wh{\beta}_{j_0}$ for $\beta_{j_0}^*$ and its plug-in variance estimator $\wh{\sigma}_{j_0}^2$ for $\sigma_{j_0}^2$ obtained from Algorithm~\ref{algo:psi}. Under Assumptions~\ref{assume:submodel:covariate} and \ref{assume:submodel:selection}, we have $\mathbb{E}(\wh{\beta}_{j_0}) - \beta_{j_0}^* = o(n^{-3/4})$. Additionally under Assumption~\ref{assume:submodel:covariate2}, we have $\mathbb{E}(\wh{\sigma}_{j_0}^2) - \sigma_{j_0}^2 = o(n^{-1/2})$.
\end{Theorem}

Theorem~\ref{result:psi:bias} provides the bias orders of the SILAB estimator $\wh{\beta}_{j_0}$ and its plug-in variance estimator $\wh{\sigma}_{j_0}^2$. As demonstrated in Section~\ref{sec:theory_submodel}, the small bias orders of these estimators can lead to accurate inference performance in finite samples, as demonstrated in our numerical studies in Sections~\ref{sec:simulations} and \ref{sec:case_study}.

\section{Simulations}
\label{sec:simulations}

In this section, we examine the finite sample performance of our proposed SILAB method by comparing it to some existing inference methods for high-dimensional logistic regression models. We emphasize the reduced bias of the SILAB estimator and demonstrate its advantages in improving the inference accuracy. To compare the inference performance, we focus on one-sided problems, where the bias tends to have a larger influence on the coverage probability of the CI compared to two-sided problems. Moreover, since it is difficult to compare one-sided CIs (e.g., in terms of CI lengths), we formulate them into one-sided tests and compare different methods in terms of the type-I error and the statistical power.

Specifically, we set $n=400$ and consider a full logistic regression model $\cS_F$ with dimension $d\in\{400,800,1600\}$. For the true parameter $\bb^*$, for any $j\in [d]$ we set
\bsq
    \beta_j^* = \left\{
	\begin{array}{ll}
		0.25  & \mbox{if } j \in\{4,8,12,16,20\} \\
            \frac{3}{4\sqrt{d_0/5-1}}  & \mbox{if } j\in\{24,28,\ldots,4d_0\}  \\
		0 & \mbox{otherwise}
	\end{array},
\right. 
\esq
where we recall that $d_0$ denotes the sparsity level. In this simulation study, we consider the following two scenarios: 
\begin{enumerate}[label=\textbf{Setting \Alph*:}, wide]
    \item Fix $d_0=20$. The covariates $\{\x_i\}_{i\in[n]}$ are generated from a $\mathcal{N}(\0, \bSig)$, where $\Sigma_{kl} = \rho^{|k-l|}$ with a varying dependence level $\rho\in\{0,0.2,0.4\}$.  
    \item Let the covariates $\{\x_i\}_{i\in[n]}$ be generated from a $\mathcal{N}(\0, \bSig)$ with $\Sigma_{kl} = 0.4^{|k-l|}$. We vary the sparsity level $d_0\in\{20,30,40\}$. 
\end{enumerate}
We consider our component of interest to be $\beta_4^*$ with a true value of $0.25$. For this component of $\bb^*$, we consider the following one-sided tests with the theoretical size $\alpha$ ranging from $0.01\%$ to $10\%$. 
\begin{enumerate}[label=\textbf{Test \arabic*:}, wide]
    \item $\text{H}_0: \beta_4^* = 0.25$ vs $\text{H}_1: \beta_4^* > 0.25$, which allows to evaluate the type-I error. 
    \item $\text{H}_0: \beta_4^* = 0$ vs $\text{H}_1: \beta_4^* >0$, which allows to evaluate the statistical power.
\end{enumerate}

We compare the finite sample performance of the SILAB method to the following existing methods: (i) the Linearization with Variance Enhancement (LiVE) method proposed in \cite{guo2021inference} and implemented in the R package \texttt{SIHR} \citep{rakshit2021sihr}; (ii) the Desparsified Lasso (D-Lasso) method proposed in \cite{van2014asymptotically} and implemented in the R package \texttt{hdi} \citep{dezeure2015high}. We consider $10^4$ Monte-Carlo replications for comparison.

\begin{figure}[!tb]
    \centering
    \includegraphics[width = 0.94\textwidth]{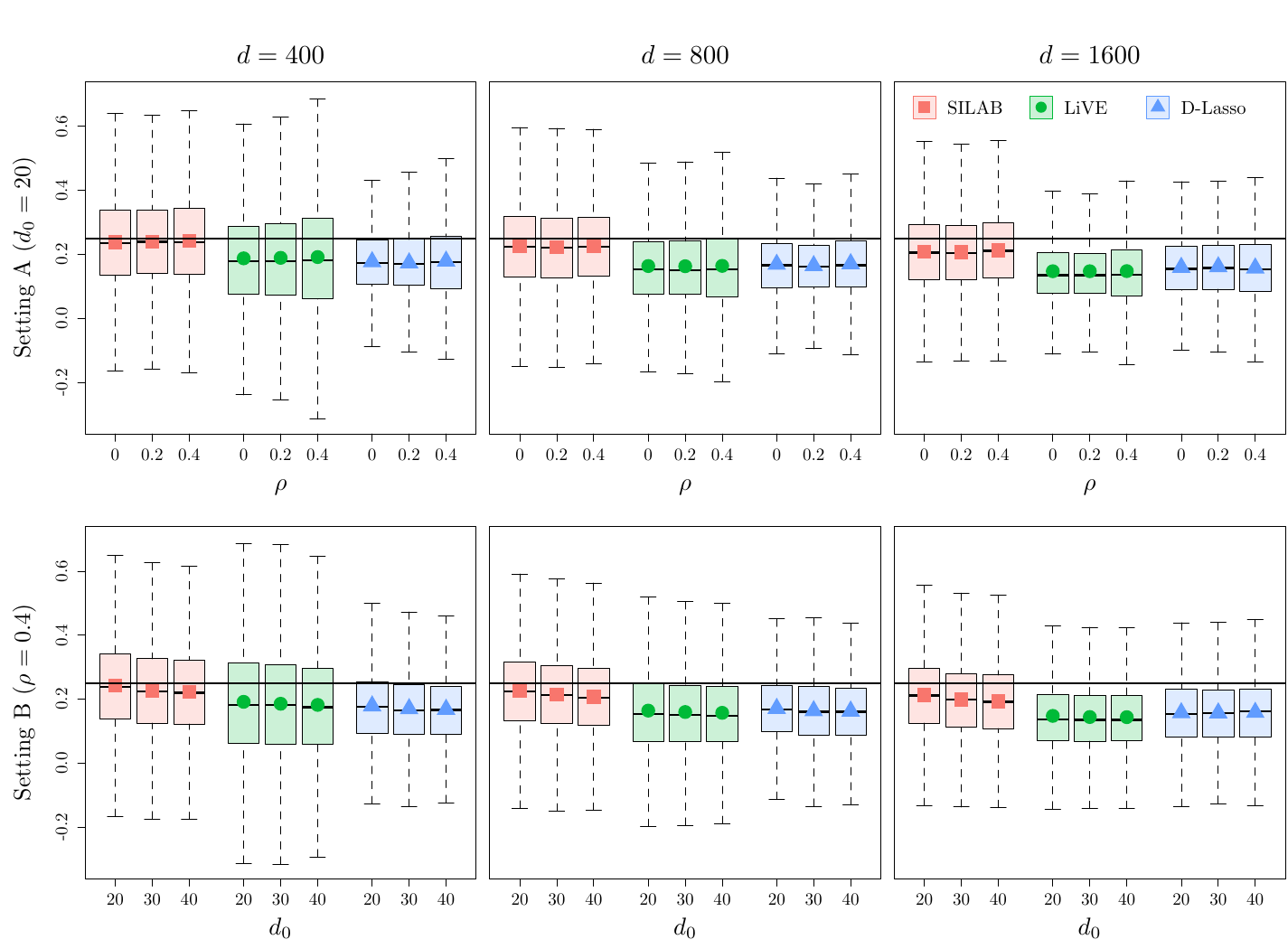}
    \caption{Empirical distributions of the point estimators for $\beta_4^*$ based on different methods. The dots represent the means of the point estimation. The horizontal lines represent the true value $0.25$ for $\beta_4^*$.}  
    \label{fig:boxplots_big_p}
\end{figure}

In Figure~\ref{fig:boxplots_big_p}, we present the empirical distributions of the point estimators for $\beta_4^*$ based on the SILAB, LiVE and D-Lasso methods. For Setting~A (see upper panel), we first observe that the SILAB estimator has the smallest and almost negligible biases, even when $d$ is relatively large compared to $n$, whereas the LiVE and D-Lasso estimators are more biased even with the smallest $d$. Moreover, as the covariates become more dependent (i.e., larger $\rho$), both the LiVE and D-Lasso estimators become more variant, whereas the variance of the SILAB estimator remains relatively stable. For Setting~B (see lower panel), we also find that the SILAB estimator is generally less biased than the LiVE and D-Lasso estimators, even in the setting with the largest $d$. In addition, as the sparsity level $d_0$ increases, all estimators have relatively stable variance. Furthermore, the variance of the SILAB estimator is generally comparable (or even smaller) than the ones of the other estimators. Therefore, these results show that the SILAB estimator is generally less biased than the LiVE and D-Lasso estimators without inflating the variance in finite samples.

\begin{figure}[!tb]
    \centering
    \includegraphics[width = 0.9\textwidth]{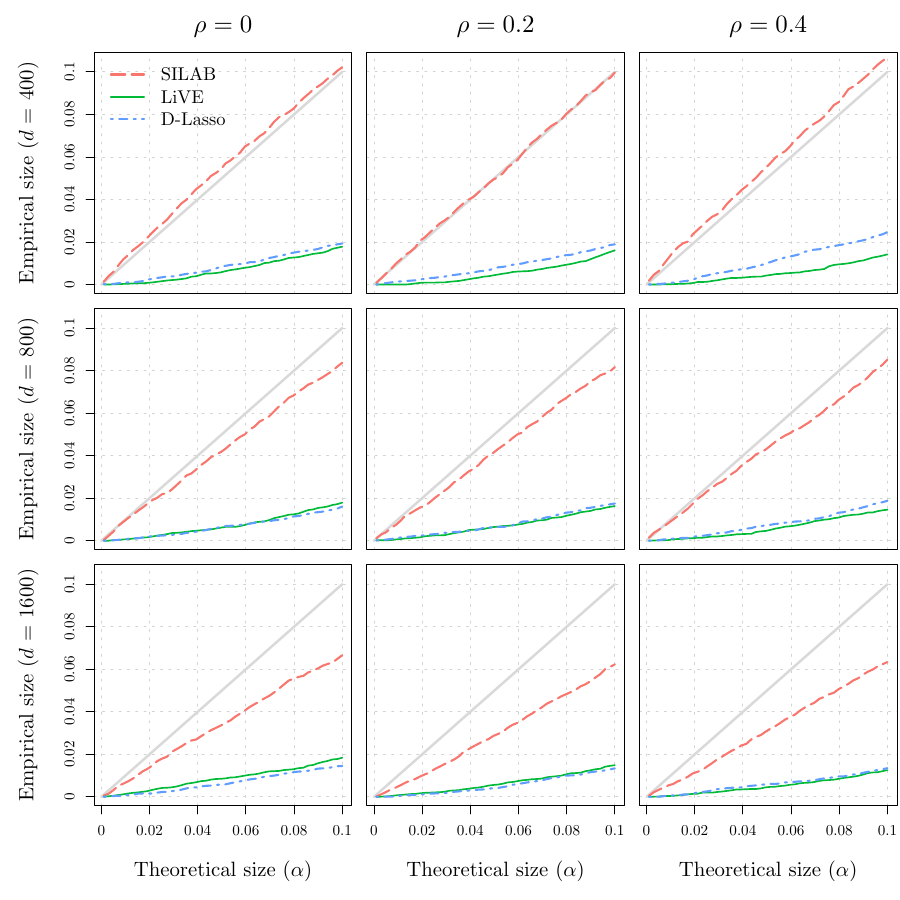}
    \caption{Type-I errors of the hypothesis test $\text{H}_0:\beta_4^*=0.25$ vs $\text{H}_1:\beta_4^*>0.25$ (i.e., Test~1) in Setting~A based on different methods. The diagonal grey lines represent the $45$ degree lines.}  
    \label{fig:sizeA}
\end{figure}

\begin{figure}[!tb]
    \centering
    \includegraphics[width = 0.9\textwidth]{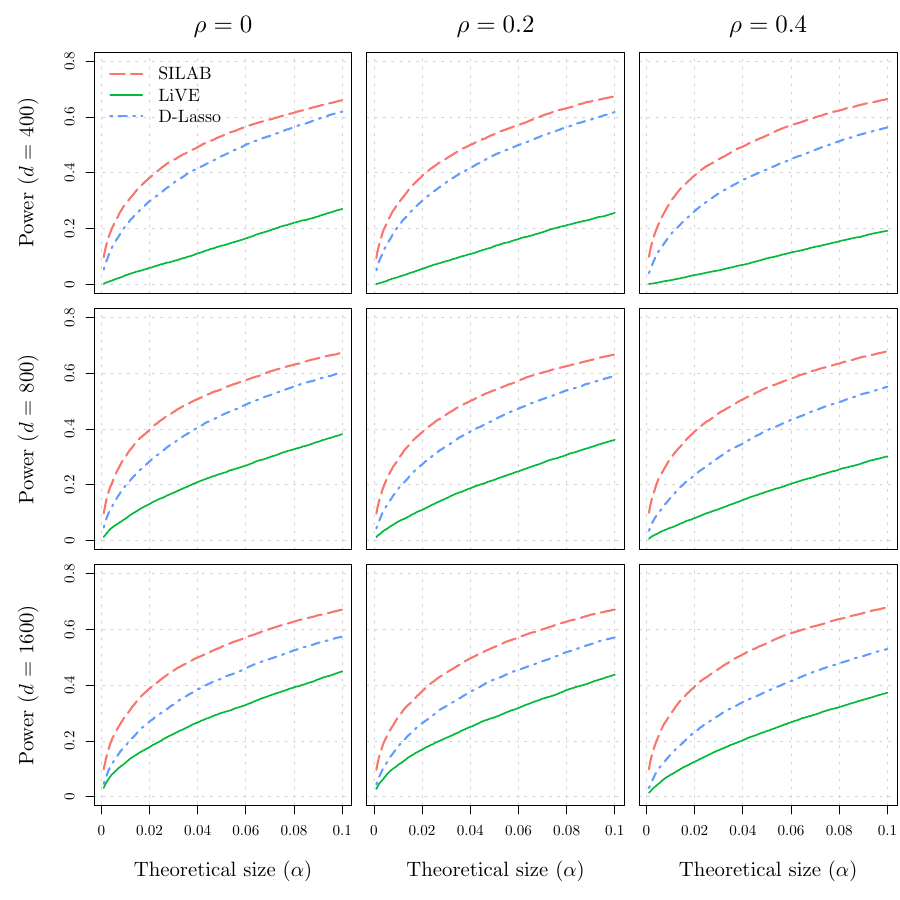}
    \caption{Statistical powers of the hypothesis test $\text{H}_0:\beta_4^*=0$ vs $\text{H}_1:\beta_4^*>0$ (i.e., Test~2) in Setting~A based on different methods.}  
    \label{fig:powerA}
\end{figure}

For the finite sample inference performance for $\beta_4^*$ in Setting~A, we present the empirical type-I error of Test~1 in Figure~\ref{fig:sizeA} and the statistical power of Test~2 in Figure~\ref{fig:powerA}. From Figure~\ref{fig:sizeA}, we observe that in general the SILAB method leads to empirical type-I errors that are significantly closer to the nominal levels than the LiVE and D-Lasso methods. Nevertheless, with the increasing dependence among covariates as well as the increase of the model dimension $d$, the performance of the SILAB method appears to deteriorate with less well controlled type-I errors, but still significantly outperforms the other methods. From Figure~\ref{fig:powerA}, we can see that the SILAB method can provide significantly more powerful tests than alternative methods, with the D-Lasso method outperforming the LiVE method. The powers of the SILAB method maintain stable with the increasing dependence among covariates or the increase of $d$, whereas the powers of the alternative methods deteriorate. For reason of space, we present the results for Setting~B in Supplementary Materials, which lead to similar observations in terms of the estimation and inference performance of all methods as in Setting~A. Overall, our numerical experiments indicate that the SILAB method can provide a powerful testing procedure with an accurate type-I error, which results from its significantly reduced bias in finite samples.

\section{Alcohol consumption data analysis}
\label{sec:case_study}

In this section, we use a logistic regression to analyze a public health real dataset from \cite{Cortez2008UsingDM} on alcohol consumption among secondary school students (15 to 22 years old) in Portugal. In particular, this dataset has $n=395$ samples and $46$ attributes. Since the two attributes of workday and weekend alcohol consumptions take five scales (i.e., very low, low, medium, high, very high), we combine them to form a binary variable to denote the alcohol consumption level, whose value is $0$ only if the workday alcohol consumption is very low and the weekend alcohol consumption is very low or low. The remaining $44$ attributes include gender, family size, quality of family relationships, number of school absences and so on. For this analysis, we consider a logistic regression model where gender and family size are included into the model only with marginal effects, whereas the remaining attributes are included up to second-order interactions. This leads to a final model with dimension $d=905$. The aim of this study is to determine the effects of gender and family size on alcohol consumption levels of secondary school students. In particular, we are interested in testing the following: 
\begin{enumerate}[label=\textbf{Test \arabic*:}, wide]
    \item Whether male has higher alcohol consumption levels than female, i.e., $\text{H}_0: \beta_{\text{gender}} = 0$ vs $\text{H}_1: \beta_{\text{gender}} > 0$.
    \item Whether students from a small family (no more than $3$ members) has higher alcohol consumption levels than the ones from a larger family, i.e., $\text{H}_0: \beta_{\text{size}} = 0$ vs $\text{H}_1: \beta_{\text{size}} > 0$.
\end{enumerate}

\begin{figure}[!tb]
    \centering
    \includegraphics[width = 0.94\textwidth]{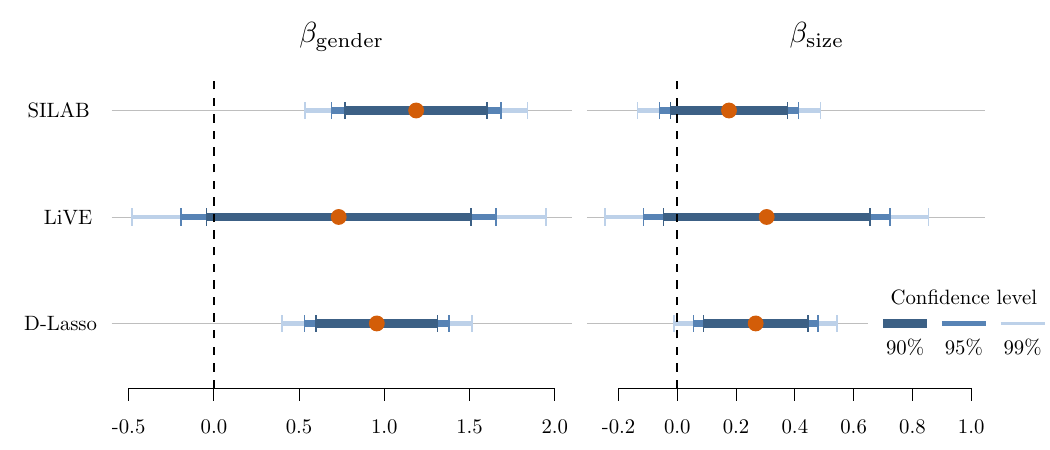}
    \caption{Point estimates (the dots) and two-sided CIs of 90\%, 95\% 99\% confidence levels for $\beta_{\text{gender}}$ and $\beta_{\text{size}}$ in the high-dimensional logistic regression model considered in Section~\ref{sec:case_study}.}  
    \label{fig:student_ci}
\end{figure}

We again compare the performance of our proposed SILAB method to the LiVE and D-Lasso methods. We illustrate the point estimates and the two-sided CIs of 90\%, 95\% 99\% confidence levels for $\beta_{\text{gender}}$ and $\beta_{\text{size}}$ in Figure~\ref{fig:student_ci}. Based on these results, in order to perform Test~1 and Test~2 at the 5\% significance level, the corresponding 95\% upper one-sided CIs are no less than the lower bounds of 90\% two-sided CIs under the null hypotheses. We can observe that, at the 5\% significance level, the SILAB method rejects the null hypothesis of Test~1, suggesting a significant evidence that male drinks more alcohol than female. Moreover, the SILAB method fails to reject the null hypothesis of Test~2, suggesting that we cannot conclude $\beta_{\text{size}} > 0$. On the other hand, the LiVE method fails to reject both null hypotheses whereas the D-Lasso method rejects both. Therefore, we reach different conclusions based on different methods. 

In the literature of alcohol use among adolescents in late adolescence and young adulthood, it has been found, for example, that boys are more at risk for problematic alcohol drinking than girls especially when entering young adulthood (see e.g., \citealt{young2002substance,schulte2009gender}). Families also have significant influences on the alcohol use of adolescents, but primarily through family quality rather than family size. For example, some research works indicate that adolescents who experience positive family bonds and reside in intact families tend to consume less alcohol. Moreover, those living with both biological parents are less likely to engage in heavy alcohol use compared to those in alternative family arrangements, such as single-parent households (see e.g., \citealt{bahr1995family,bjarnason2003alcohol}). Therefore, the inference results based on the SILAB method appear to be the most in line with the existing experimental findings, suggesting the usefulness of our proposed method.

\section{Discussion}
\label{sec:discussion}

In this paper, we put forward a two-step procedure, the SILAB method, to perform inference for low-dimensional components of the parameter in a high-dimensional logistic regression model. In the first step, we propose a novel SILA method to find a suitable submodel, which not only enjoys the sure screening property (i.e., contains the true model with probability approaching one with a moderate model size), but also reduces the influence of irrelevant variables with high spurious correlations. In the second step, we propose to fit the BC-MLE on the submodel from the first step for further inference. Under a set of easily verifiable conditions on the covariates, we show that the BC-MLE enjoys a higher-order bias correction property, which yields improvement in the accuracy of subsequent inference, especially for one-sided CIs and tests that are commonly considered in various disciplines. 

Many extensions are worth investigating following our work. For example, we mention that the key for our proposed method to perform well is that the submodel selected in the first step does not underfit and tends to overfit. We choose to use the Lasso in the SILA method for variable selection owing to its tendency of overfitting. A potential extension is to incorporate this ``split-and-intersection'' idea to other methods with folded-concave penalties such as the Smoothly Clipped Absolute Deviation (SCAD, \citealt{fan2001variable}). Another possible methodological extension is to go beyond the logistic regression models to other generalized linear models, other parametric models, or even semi-parametric models.

\newpage
\normalem 
\bibliographystyle{apalike}
\bibliography{bibliography}
\end{document}